\documentclass[journal, draftclsnofoot,onecolumn,12pt]{IEEEtran}
\usepackage{cite}
\usepackage{bm}
\usepackage{siunitx}
\ifCLASSINFOpdf
\else
\fi
\usepackage{amsmath}
\usepackage{ amssymb }
\usepackage{amsfonts}
\usepackage{caption}
\usepackage{graphicx}
\usepackage{lettrine}
\usepackage{epstopdf}
\usepackage{textcomp,booktabs}
\usepackage[usenames,dvipsnames]{color}
\usepackage{colortbl}
\definecolor{mygray}{gray}{.9}
\definecolor{mypink}{rgb}{.99,.91,.95}
\definecolor{mycyan}{cmyk}{.3,0,0,0}
\usepackage{pifont}
\usepackage{subfig}
\usepackage{multirow}
\usepackage{url}
\usepackage{color}

\usepackage{setspace}
\usepackage{lineno}

\usepackage[dvipsnames,usenames]{color}
\usepackage[usenames,dvipsnames]{color}

\definecolor{light-gray}{gray}{0.90}

\begin{document}
	\title{AI-aided Online Adaptive OFDM Receiver:  Design and Experimental Results}

	
	\author{Peiwen Jiang, Tianqi Wang, Bin Han, Xuanxuan Gao, Jing Zhang, \\ Chao-Kai Wen, Shi Jin, and Geoffrey Ye Li
			\thanks{P. Jiang, T. Wang, B. Han, X. Gao, J. Zhang and S. Jin are with the National
				Mobile Communications Research Laboratory, Southeast University, Nanjing
				210096, China (e-mail: PeiwenJiang@seu.edu.cn; wangtianqi@seu.edu.cn; seuhanbin@seu.edu.cn;  gaoxuanxuan@seu.edu.cn; jingzhang@seu.edu.cn; jinshi@seu.edu.cn).}
			\thanks{C.-K. Wen is with the Institute of Communications Engineering, National
				Sun Yat-sen University, Kaohsiung 80424, Taiwan (e-mail: chaokai.wen@mail.nsysu.edu.tw).}
			\thanks{G. Y. Li is with the Department of Electrical and Electronic Engineering,
				Imperial Colledge London, London, UK (e-mail: geoffrey.li@imperial.ac.uk).}}
	
	\maketitle
%
%
%
\begin{abstract}
		
	Orthogonal frequency division multiplexing (OFDM) has been widely applied in current communication systems.  The artificial intelligence (AI)-aided OFDM receivers are currently brought to the forefront to replace and improve the traditional OFDM receivers. In this study, we first compare two AI-aided OFDM receivers, namely, data-driven fully connected deep neural network  and model-driven ComNet, through extensive simulation and  real-time video transmission using a 5G rapid prototyping  system for an over-the-air (OTA) test. We find a performance gap between the simulation and the OTA test caused by the discrepancy between the channel model for offline training and the real environment. We develop a novel online training system, which is called SwitchNet receiver, to address this issue. This receiver has a flexible and extendable architecture and can adapt to real channels by training  only several parameters online. From the OTA test, the AI-aided OFDM receivers, especially the SwitchNet receiver, are robust to real environments and  promising for future communication systems. We discuss potential challenges and future research inspired by our initial study in this paper.
	\end{abstract}
	
	
	\newpage
	
	\setlength{\baselineskip}{22pt}
	\section{Introduction}

	\IEEEPARstart{O}{rthogonal} frequency division multiplexing (OFDM) is an effective technique to deal with the delay spread of wireless channels \cite{701317,cho2010mimo}.
	The conventional OFDM receivers can be classified into two categories: linear and iterative receivers. Linear receivers include least square (LS) \cite{coleri2002channel, simeone2004pilot} and minimum mean-squared error (MMSE) \cite{Myllyla2005ComplexityAO} for channel estimation (CE) or signal detection (SD). Iterative receivers include approximate message passing  \cite{rangan2011generalized} and expectation propagation-based algorithms \cite{wu2016block}. These receivers are all designed  on the basis of expert knowledge or specific models. However, complex channel scenarios and nonlinear interference challenge these conventional designs and limit the performance of OFDM receivers.	
	 OFDM can potentially address many challenging issues in traditional  systems with the introduction of artificial intelligence (AI).  \cite{8054694, qin2018Deep,DBLP:journals/corr/abs-1809-06059,2019arXiv190606007G} revealed the benefits of AI in physical layer of communications. Different modules in the conventional communication systems have been studied with the aid of AI, including signal classification \cite{o2018over},  channel coding \cite{gruber2017deep,nachmani2017rnn,liang2017iterative}, CE \cite{soltani2019deep}, multiple-input multiple-output (MIMO) detection \cite{DBLP:journals/corr/abs-1809-09336,samuel2017deep}, channel state information (CSI) feedback \cite{8322184, 8482358}, and novel autoencoder-based end-to-end communication systems \cite{D2018Deep, ye2018channel}.	
	
	 A novel data-driven AI-aided OFDM receiver is recently proposed in \cite{8052521}, and this receiver uses a fully connected deep neural network (FC-DNN) to detect data directly without estimating CSI explicitly after applying a fast Fourier transformation (FFT) module.  The AI-aided OFDM receiver, which treats joint CE and SD as a black box, exploits no expert knowledge of wireless communications and trains all parameters with a large amount of wireless data by stochastic gradient descent-based algorithms. The data-driven AI-aided OFDM receiver in \cite{8052521} is robust to the effect of pilot reduction, CP omission, and nonlinear clipping noise; however, it converges slowly and has high computational complexity.   A deep complex convolutional network (DCCN), which is inspired by \cite{8052521}, is proposed in \cite{OFDM2018DL1} for the receiver to convert OFDM waveform into detected symbols directly without using discrete Fourier transform (DFT). The Cascade-Net in \cite{OFDM2018DL2} is robust to ill-condition channels. The deep learning (DL) networks in \cite{balevi2019one} can perform CE and data symbol detection for one-bit OFDM receivers. Many other data-driven methods, such as those in \cite{gui2018deep,xu2018deep}, are also  developed recently.   In brief, the AI-aided OFDM receiver does not need to know prior information on the hand-craft and usually outperforms traditional OFDM receivers in terms of BER performance.
	
	AI algorithms can exploit expert knowledge to develop model-driven AI approaches. One of the earliest model-driven AI approaches in \cite{sun2016deep} is proposed for magnetic resonance imaging.  Model-driven AI approaches are currently extended to wireless physical layers through the design of network architectures based on wireless physical domain knowledge\cite{DBLP:journals/corr/abs-1809-06059}; they are promising in addressing  the CE and SD problems. In particular, a model-driven-based AI-aided OFDM receiver, which is  called ComNet, is proposed in\cite{gao2018comnet}. Instead of using a single DNN to detect signals with implicit CE as in the FC-DNN receiver \cite{8052521}, ComNet follows the conventional OFDM architecture but uses two DNNs for CE and SD, respectively, to improve the performance of the modules. Simulation results showed that ComNet has better performance than the traditional MMSE-based methods and converges faster given that only fewer parameters need to be trained than the FC-DNN OFDM receiver \cite{8052521}. Furthermore, explicit CE helps in channel analysis and CSI feedback in downlink transmission, especially in massive MIMO systems. The abovementioned advantages render ComNet a competitive candidate for practical system implementation. Additional research in this topic can be found in \cite{8240644,8445920}.
	
	Although the abovementioned AI-aided methods work well based on simulation, the over-the-air (OTA) performance in practical environments remains unknown. State-of-the-art OTA studies usually train well-designed AI networks offline and deploy them on software-defined radios (SDRs), such as universal software radio peripheral (USRP), for online use \cite{o2018over,D2018Deep}. In this case, the trained parameters of the DNNs remain the same as they are deployed. Therefore, all possible effects of practical environments have to be considered during the architecture design and training phase, which is  impractical in some cases. In \cite{8491189},   error-correcting codes (ECCs) are used to construct labeled datasets at the receiver side such that the trained AI communication systems can be fine-tuned by transfer learning at run time. This method only works for channels varying  slower than updating parameters. Apart from transfer learning,  the low-complex frameworks in \cite{liu2019online} and the meta-learning frameworks in \cite{mao2019roemnet,park2020end} significantly reduce required training resources for online adaptation. To the best of our knowledge,  report on using AI-aided OFDM receivers in real environments with a real-time video transmission is lacking. Many practical issues are  challenging, such as the robustness to the varying channels, the choice of the training data, and the real-time data processing. 
	
	In this work, we compare the FC-DNN OFDM receiver \cite{8052521} and the ComNet OFDM receiver \cite{gao2018comnet} through an OTA test because many details may be ignored in simulation.
We set up a real-time video transmission system based on the two AI receivers for the OTA test by using the 5G rapid prototyping (RaPro) system in \cite{yang2017rapro, Gao2018Implementation}. The OTA test in diverse environments demonstrates that the AI-aided OFDM receivers are feasible and extendable in practical applications.
However, we find a performance gap between the simulation and the OTA test.
We develop an online learning architecture, which is called SwitchNet receiver and can be trained with offline and real-time online data, to capture channel features ignored during offline training for addressing the abovementioned problem.    Compared with current AI-aided OFDM receivers, the proposed SwitchNet has a novel architecture that can be trained online and can still recover the transmitted symbols even when the training channel data set and the practical channel conditions mismatch. In addition, the number of online trainable parameters is small, which can reduce the online training data and cost. Furthermore, the proposed SwitchNet is implemented and evaluated OTA, which proves the feasibility and the real-time transmission ability of the online training method in communication. 
  The contributions of this study are summarized as follows:
		
		1) We   comprehensively compare the performance of the current AI-based receivers, including the FC-DNN, ComNet, and the conventional LMMSE receiver. We  reveal the sensitivity of the AI-based receivers to the mismatch of channels between offline training  and online deployment stages; this condition indicates the necessity of online training schemes.

		2) We propose a novel online training method, which is called SwitchNet. This method investigates multiple channel models during offline training and leaves only a few trainable parameters to be tuned by online training. The network   can be applied to a wide range of environments even without re-training. Furthermore,  the network can converge   very fast to adapt to real environments and avoid overfitting by online training.
		
		3) We demonstrate the feasibility of online training SwitchNet in different scenarios through a real-time video transmission rapid prototyping (RaPro) system as the OTA test platform to deploy AI OFDM receivers. To the best of our knowledge, it is one of the few online training AI-aided OFDM communication systems.

	The rest of this paper is organized as follows. Section \uppercase\expandafter{\romannumeral2} introduces the architectures of the FC-DNN receiver, ComNet receiver, and SwitchNet. Simulation results are shown and discussed in Section \uppercase\expandafter{\romannumeral3}. In Section \uppercase\expandafter{\romannumeral4}, we analyze the OTA test results. We summarize the challenges for future work in Section \uppercase\expandafter{\romannumeral5}.
	
\setlength{\baselineskip}{20pt}	
	\section{Architectures of AI-aided OFDM receivers}
	In this section, the traditional and AI-aided OFDM systems are introduced.  After  the existing data-driven FC-DNN receiver \cite{8052521} and the model-driven ComNet receiver  \cite{gao2018comnet} are analyzed, we develop SwitchNet to facilitate an OTA test and the practical application of the AI-aided OFDM receiver.
	
	\begin{figure*}[t]
		\centering
		\includegraphics[width=6.8in]{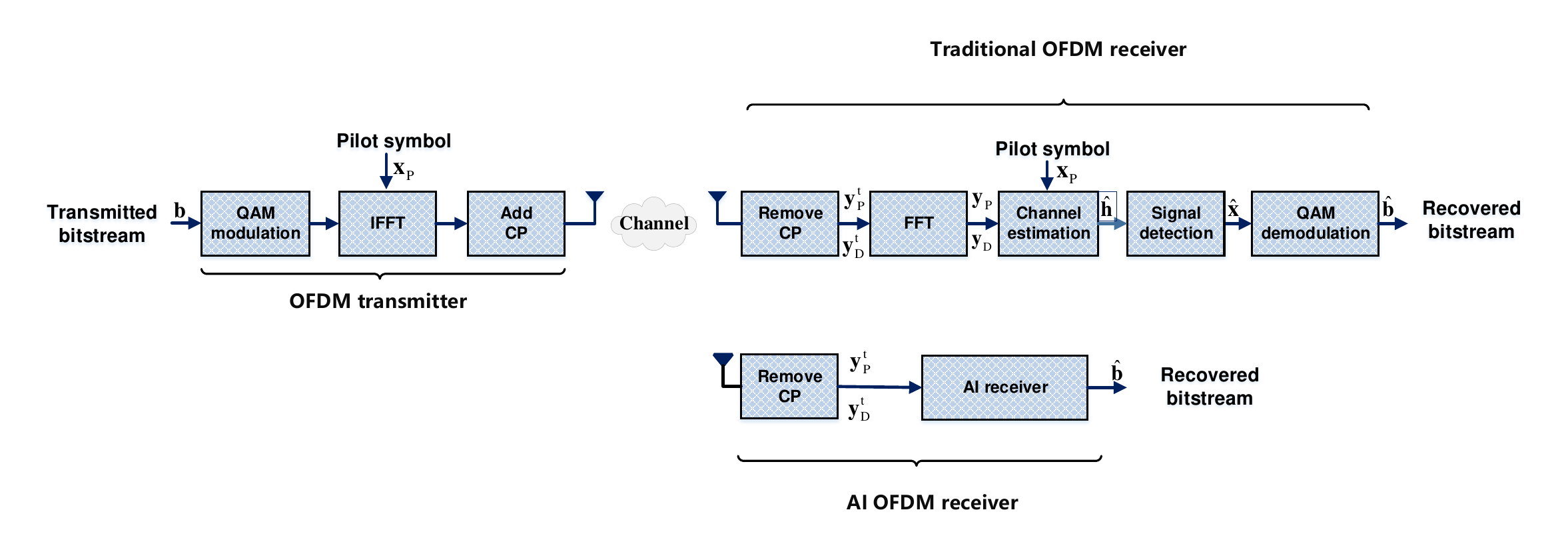}
		\caption{Block diagram of an OFDM system, where pilot symbols are inserted at the transmitter, and the receiver acquires CSI. Compared with the traditional OFDM receiver, the AI receiver performs CE, SD, and QAM demodulation altogether and directly maps the received signals into decided symbols.}
		\label{fig_1}
	\end{figure*}
	
	\subsection{Traditional and AI-aided OFDM systems}
	Fig. \ref{fig_1} shows the block diagram of an OFDM system, including transmitter, channel, and receiver. Two types of OFDM receivers are introduced: traditional and AI-aided OFDM receivers.

 For the transmitter, the input binary data sequence, $\mathbf{b}$, is modulated as the transmit symbol sequence, where $M$-QAM, such as 4-QAM and 16-QAM, is used. Then, an $N$-point IFFT is performed to generate an OFDM signal. Thereafter, a cyclic prefix (CP) is inserted to mitigate the inter-symbol interference (ISI) caused by the delay spread of wireless channels.

 Wireless channels are assumed to be with delay spread and additive white Gaussian noise, $\mathbf{w}$. The components of $\mathbf{w}$ are independent, with zero-mean and $\sigma _{w }^{2}$-variance. A sample-spaced multipath channel can be described by complex random variables. Without the CP, the delay spread of $L-1$ samples will result in ISI and inter-carrier interference (ICI).  If the delay spread is shorter than the length of the CP, $P$, that is, $L-1\le P$, then   ISI and ICI will not occur. To estimate CSI, pilot symbols are inserted in the first OFDM block in a frame while the transmitted data are appended in the following OFDM blocks of the frame. The channel is assumed to be time-invariant during one frame.
	
	At the traditional OFDM receiver in Fig. \ref{fig_1}, the CP is removed first and then FFT is performed. We denote $y(k)$ as the received signal at the $k$-th subcarrier. CE, SD, and QAM demodulation are subsequently performed. The received pilot and data signals for the $k$th subcarrier can be expressed as
	\begin{align*}
	& {{y}_{P}}(k)={{x}_{P}}(k)h(k)+w(k),
	\end{align*}
	and
	\begin{align*}
	& {{y}_{D}}(k)={{x}_{D}}(k)h(k)+w(k),
	\end{align*}
	respectively, where ${x}_{P}(k)$ and ${x}_{D}(k)$ denote the pilot and transmit symbols in the $k$th subcarrier, respectively. ${x}_{P}(k)$ is known at the receiver and  used for  CE while ${x}_{D}(k)$ is unknown at the receiver and needs to be detected based on the received signal and estimated channel.
	
	The AI receiver in Fig. \ref{fig_1} replaces the three latter modules in the traditional receiver. It directly maps the received symbols into the detected binary data. In the following sections, two types of AI receivers, namely, data-driven FC-DNN and the model-driven ComNet, and a novel AI receiver, called SwitchNet are described in detail.

	\subsection{FC-DNN receiver}
\begin{figure}[!t]
		\centering
		\includegraphics[width=3.3in]{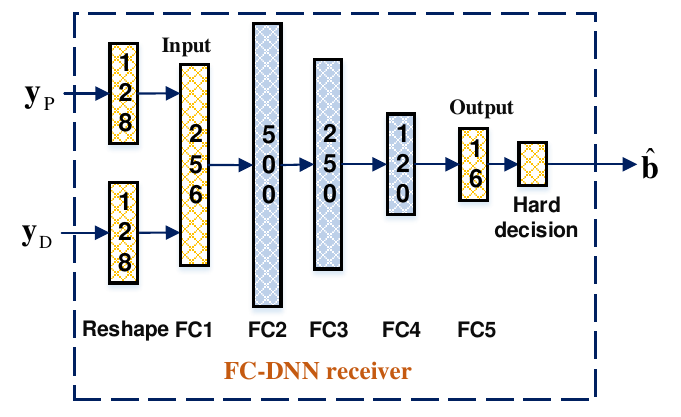}
		\caption{Structure of FC-DNN. The FC-DNN receiver contains five FC layers that directly map the received signal to the recovered bitstreams.}
		\label{fig_2}
	\end{figure}

	A data-driven AI-aided FC-DNN receiver in Fig. \ref{fig_2} is proposed in \cite{8052521}. The received signals, including pilot and data, are reshaped as the input from complex value to real value initially.  Therefore, the size of input layer will be $2(K_{P}+K_{D})$ if  $K_{P}$ subcarrier pilot and $K_{D}$ subcarrier data are available.  Then, the input data go through three hidden layers.    The numbers of {neurons} in the hidden layers are 500, 250, and 120.  The output layer is composed of only ${N}/{\text{8}}$ neurons and 8 FC-DNN receivers are designed to work in parallel to recover all input binary data,   where $N$ is the length of the input binary data. Such design is beneficial for ensuring high precision of the estimated symbols and avoiding a more complex network.  All but the output of layers use the ReLU function $f_{\rm Re}(a)=\max ({0,a})$ as the activation function. The activation function of the output layer is the logistic sigmoid function $f_{\rm Si}(a)=\frac{1}{{1 + {e^{ - a}}}}$ for classification. The logistic sigmoid function at the output layer maps the input to the interval $[0,1]$, which can be regarded as a soft decision. Hard decisions can be obtained on the basis of the soft decisions. Given that each DNN recovers only one-eighth of the transmit data, eight identical DNNs with different coefficients are needed to recover all the transmit binary data.
	%
	
The FC-DNN receiver merges CE, SD, and OFDM modulation into one black box and exploits offline training but an online deployment method. In the training stage, transmit bits are generated randomly as a label and then modulated to form a frame by inserting pilot symbols. The CSI is simulated by a specific channel model and varies with each frame. The ${{\ell }_{2}}$ loss and the adaptive moment estimator (Adam) optimizer \cite{Kingma2014Adam} are used in the training process.

The FC-DNN OFDM receiver adopts an end-to-end structure to realize  global optimization of the receiver. It is robust to nonlinear distortions and potential hardware imperfections, such as lack of CP and clipping. However, FC-DNN requires a large labeled data set to train its weights and converges slowly because of the large number of weights that need to be trained.

		\subsection{ComNet receiver}
\begin{figure}[!t]
			\centering
			\includegraphics[width=4.5in]{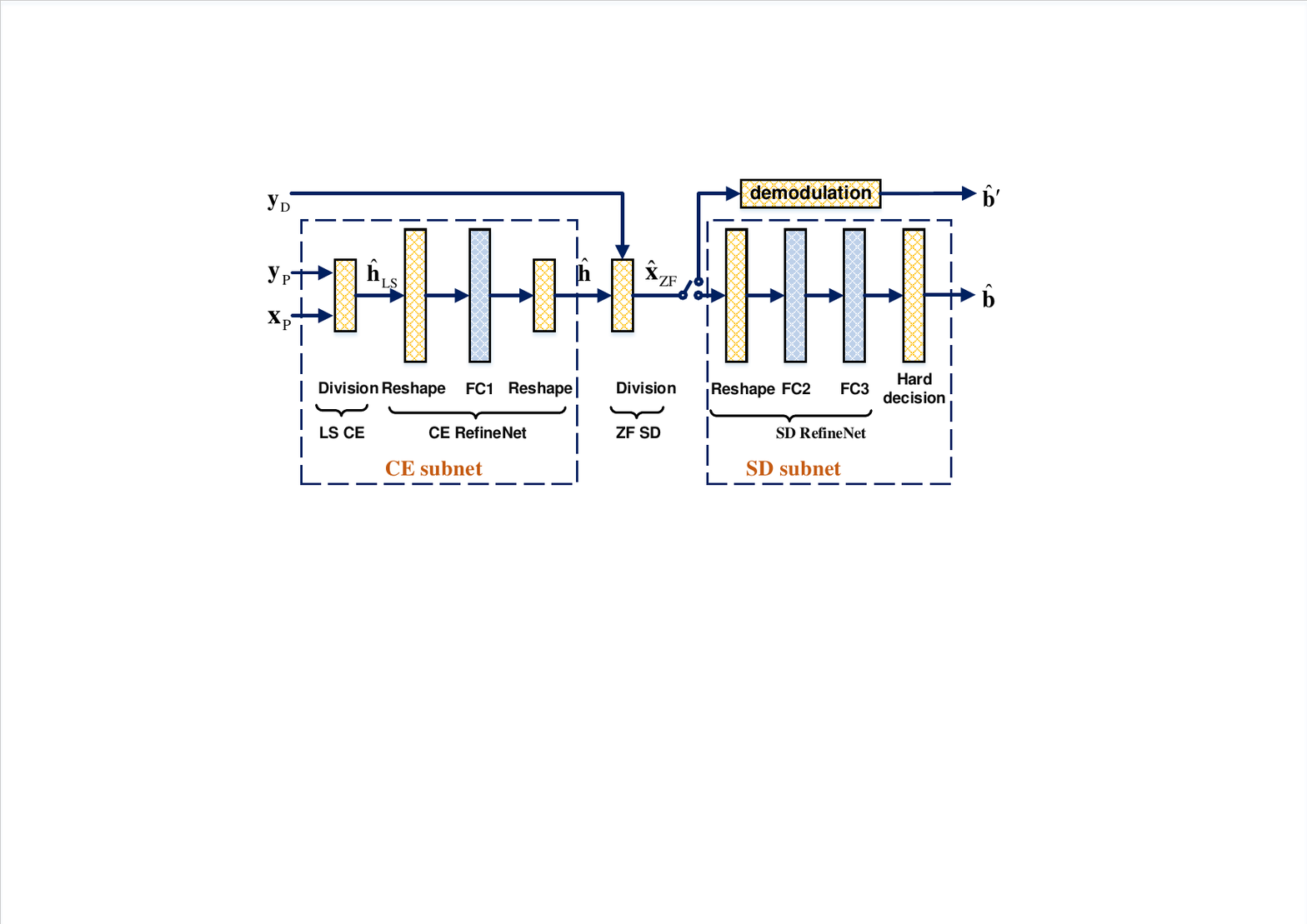}
			\caption{ComNet receiver architecture. The two subnets use traditional communication solutions as initializations and apply DL networks to refine the coarse inputs. The crossing short-path provides a relatively robust candidate of the binary symbol recovery.}
			\label{ComNet details}
		\end{figure}

		A model-driven AI-aided ComNet receiver is proposed in \cite{gao2018comnet} to alleviate the demand for vast training data and enabling the acquisition of CSI. The basic idea of the ComNet receiver \cite{gao2018comnet} is to use DNNs as auxiliary blocks to refine the original modules in the OFDM receiver in Fig. \ref{fig_1}.
		Fig. \ref{ComNet details} illustrates the architecture of the ComNet receiver  \cite{gao2018comnet}, which adopts two cascaded DNN-based subnets:  CE and SD subnets.
			
		In the CE subnet, the LS CE is  calculated by element-wise division.
 Then, the LMMSE matrix is used as the initiation.
 In the SD subnet, the zero-forcing (ZF) SD is first obtained by element-wise division.
		  The hidden layer FC2 in Fig. \ref{ComNet details} uses the ReLU activation function, whereas the output layer FC3 uses the logistic sigmoid function. A hard decision is made to decide the transmit bits. Alternatively, a short-path of a conventional QAM demodulation module can be added to obtain a robust bitstream depending on the scenario.

		Similar to the FC-DNN receiver in \cite{8052521}, the ComNet receiver \cite{gao2018comnet} uses offline training but online deployment method. Different from the FC-DNN receiver, which executes end-to-end training\cite{8052521}, the ComNet receiver \cite{gao2018comnet} adopts a two-stage training where the CE and SD subnets are trained separately and successively. Once the training process of the CE subnet is done, the parameters in the CE subnet will be fixed in the following training process of the SD subnet.  The labels of the training data include randomly generated transmitted bitstreams for updating the SD subnet and the specific channel model for updating the CE subnet. 

The  ComNet receiver \cite{gao2018comnet} exploits expert knowledge and breaks the black box of the purely data-driven AI receiver in \cite{8052521}.   Similar to the FC-DNN receiver, the output layer of the SD subnet has ${N}/{\text{8}}$ neurons and 8 SD subnets that work together in the ComNet receiver.  
	
	\setlength{\belowcaptionskip}{-0.3cm}   

	\subsection{SwitchNet receiver}
	\label{SwitchNet_receiver}
	The DNN networks in the abovementioned FC-DNN and ComNet receivers are trained with simulated data offline. This condition will lead to mismatch and performance degradation if practical channels are different from the simulated ones or unexpected distortions are ignored during offline training.

The delay spread of the multipath channel is an important parameter for calculating the LMMSE weight matrix in the CE subnet.  The robust LMMSE receiver in \cite{701317} needs the max delay to calculate the LMMSE filter matrix offline.   The delay spread estimation exploiting few channel samples online is proposed in \cite{cho2010mimo} by introducing the channel shape in time domain.  

However, AI-aided receivers learn the environmental features implicitly and the learned parameters are unexplained.  Therefore, retraining the NNs online is more difficult compared with recalculating the LMMSE matrix similar to conventional ways \cite{701317,cho2010mimo}. Therefore, an adaptive and practical AI-aided OFDM receiver is desired.  Online transmission data \cite{8491189} should also be considered in the training process of DNNs in OFDM receivers for designing a practical AI-aided OFDM receiver. 

\begin{figure}[!h]
		\centering
		\includegraphics[width=4in]{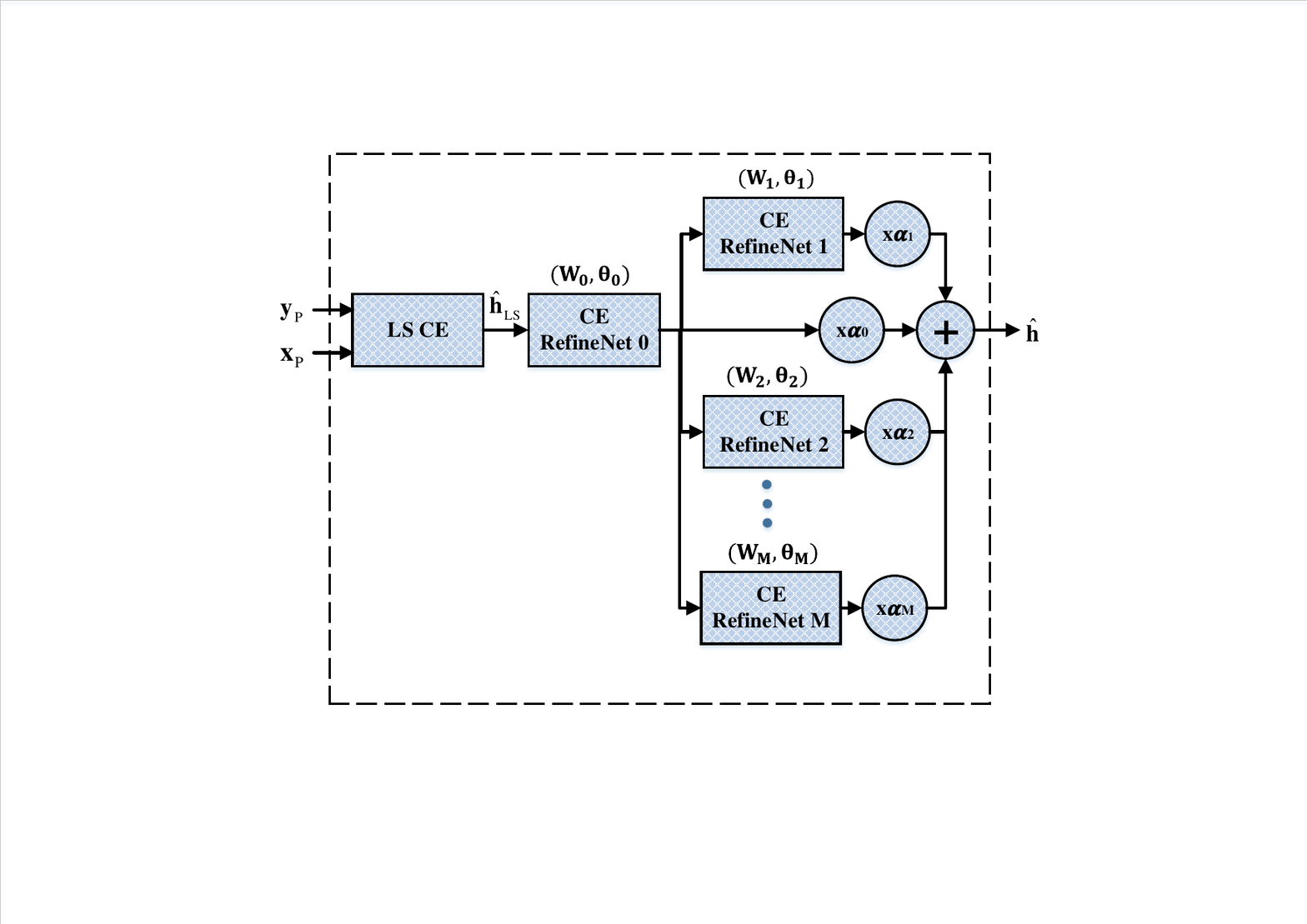}
		\caption{CE subnet architecture of the SwitchNet receiver. The CE RefineNet 0 is the basic DNN network for CE, and the CE RefineNets from 1 to $M$ are the compensating network of the CE RefineNet 0. $\alpha$ is the switch parameters to decide whether the CE RefineNets from 0 to $M$ are accessed.}
		\label{Switch}
	\end{figure}

	As shown in Fig. \ref{Switch}, the SwitchNet receiver is based on the ComNet receiver  because it is similar to conventional receivers. The difference between them is the architecture of the CE subnet is designed for online adaption. As shown in Fig. \ref{Switch}, the CE subnet of the SwitchNet receiver consists of LS CE, two or more CE RefineNets, and  online training parameters $\alpha \in [0,1]$. The structures of the LS CE and each CE RefineNet are the same as those in the ComNet receiver.   We consider two channel models, namely, the short  and  long channels, to combat the changing delay spread online. However, the architecture can be directly extended to more channel models. As depicted in Fig. \ref{Switch}, the CE RefineNet 0 is a basic neural network for channel estimation and the CE RefineNet  {from 1 to $M$} is the compensating network of the CE RefineNet 0  to adapt different channel environments.
	
 		The aforementioned CE RefineNets are trained offline for different channel models, and the switch parameters $\alpha$ are trainable online to decide whether the CE RefineNet from 0 to $M$ is accessed. Given only ${M}$ training parameters $\alpha$, a small batch of OFDM symbols with bit labels are needed and overfitting can be avoided. In the offline stage, the CE RefineNet 0 is trained for the robust  channel first, which only assumes the max delay in time domain as in \cite{701317}. Then, the trained parameters of  CE RefineNet 0 are fixed and the CE RefineNet 1 is trained to adapt the short channel.     RefineNets from 2 to $M$ have the same process for the different channels.  In the online stage,   the parameters $\alpha$ are trained to switch to the specific channel. For example, under the short channel,  all $\alpha$ connected to the RefineNets are set as 0 and only CE RefineNet 1 is accessed.   Denote ${\bf{W}}_{i}$ and ${\bm{\theta}}_{i}$ as the $2K\times 2K$ real multiplicative parameter matrix and the $2K\times 1$ real additive parameter vector for the $i$-th CE RefineNet in Fig. \ref{Switch}, where $K$ is the number of effective subcarriers.
 From Fig. \ref{Switch}, the estimated channel ${\bf \hat{h}}$ can be expressed as
	\begin{equation}
	{\bf \hat{h}}=\left(\sum_{i=1}^{M}\alpha_{i}{\bf{W}}_{i}+{ \alpha_{0}\bf I}\right)({\bf{W}}_{0}{\bf \hat{h}}_{ls}+{\bm{\theta}}_{0})+\sum_{i=1}^{M}\alpha_{i}{\bm{\theta}}_{i}.
	\end{equation}

 The offline training process is divided into two steps similar to  the ComNet. In the first step, the CE RefineNets are trained  one by one, which can be written as
	\begin{equation}
		(\hat{\mathbf{W}}_{i},\hat{\bm{\theta}}_{i})=\mathop{\arg\min}\limits_{ {\mathbf{W}}_{i},{\bm{\theta}}_{i}}|| {\bf \hat{h} - h} || ^2_2,
	\end{equation}
	where 
	\begin{equation}
		\alpha_{j}=\left\{
		\begin{aligned}
			1, & & j=i,\\
			0, & & j\neq i.\\
		\end{aligned}
		\right. 
	\end{equation}
	Then,  the CE RefineNets are fixed and the SD subnet is trained as 
	\begin{equation}
		(\hat{\mathbf{W}}_{\rm SD},\hat{\bm{\theta}}_{\rm SD})=\mathop{\arg\min}\limits_{ {\mathbf{W}}_{\rm SD},{\bm{\theta}}_{\rm SD}}|| {\bf \hat{b} - b} ||^2_2,
	\end{equation}
	where ${\mathbf{W}}_{\rm SD}$ and ${\bm{\theta}}_{\rm SD}$ are  trainable parameters in the SD subnet, $\mathbf{\hat{b}}$ is the output of the SD subnet, and $\mathbf{b}$ is the transmit bitstream.
	
	The online training strategy aims to learn a combination of the CE RefineNets with extreme few pilots and known bits. If all  bits in the data block of the subframes are known to the receiver, then they  are called  training subframes in the following.  This  end-to-end training process can be expressed as
	\begin{equation}
		\hat{\bm{\alpha}}=\mathop{\arg\min}\limits_{ \bm{\alpha}}|| {\bf \hat{b} - b} ||^2_2,
	\end{equation} 
	where $\bm{\alpha}=[\alpha_0,\alpha_1,\cdots,\alpha_M] $.  The known bits are better than ECC in online training\cite{8491189}. However, the training subframes are insufficient due to the limited transmit resources.

	The SwitchNet receiver introduces the idea of online training and can adjust to different channel environments; thus, making the OFDM system becomes more robust than the FC-DNN and the ComNet receivers.  Compared with offline switch, the online training  of SwitchNet receiver can improve the BER performance  continuously in the switch process. Meanwhile,  $\alpha$ can be beyond 0 and 1 with a good combination of the CE RefineNets to adapt a new environment. In addition, the SwitchNet receiver can be naturally extended to   multi-user MIMO systems given that the SwitchNet receiver is inherited from the ComNet, in which the traditional methods, including the LS CE and the ZF SD, can be used. 
	

	
	\section{Simulation and Discussion}
	\label{simulation}
	In this section, we analyze  the pros and cons of different AI-aided OFDM receivers through extensive simulation.
	
	\subsection{Configurations of the simulation system}
	\subsubsection{Frame structure}

\begin{figure}[!t]
		\centering
		\includegraphics[width=4in]{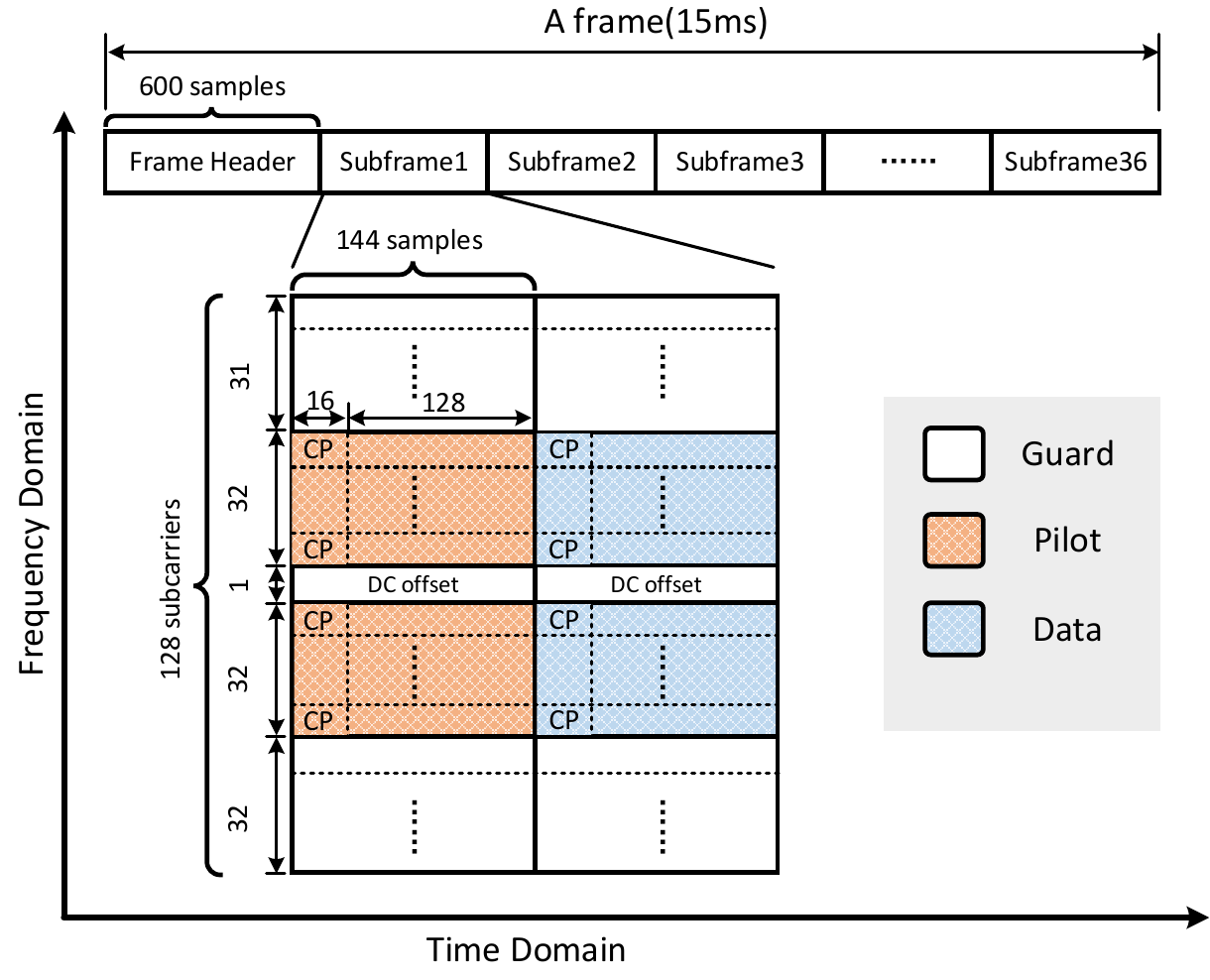}
		\caption{Frame structure of the simulated OFDM system. Each frame contains several OFDM symbols. A pilot symbol and a data symbol are set as the inputs of NNs. Each OFDM symbol contains 128  subcarriers. A total of 64 subcarriers are used for the pilot symbol or data symbol transmission while the vest 64 subcarriers serve as guard band and DC offset.}
		\label{frame}
	\end{figure}

 Fig. \ref{frame} illustrates the frame structure of the simulated OFDM system.   In Fig. \ref{frame}, a 15 ms frame is composed of a frame header and 36 subframes. Each subframe contains one pilot OFDM symbol and one data OFDM symbol.  Each OFDM symbol consists of 128  subcarriers where 64 subcarriers are used for pilot or data transmission and the others are for guard band and DC offset.

	\subsubsection{Channel conditions}
	Three  classic channel conditions are chosen for simulation to
	investigate the effect of changing scenarios on the conventional and AI-aided receivers.

	\textbf{Short channel}
	in the simulation is with the exponential (EXP) power delay profile (PDP), which is defined in IEEE 802.11b to model the indoor channel at the carrier frequency of 2.4 GHz \cite{cho2010mimo}. The PDP is given as follows:
	\begin{equation}
	P(\tau)=\frac{1}{\tau_{rms}}e^{-\tau/\tau_{rms}},
	\end{equation}
	where $P(\tau)$ is the received power at delay $\tau$, and $\tau_{rms}$ denotes the root-mean-square (RMS) delay spread.
	The output of a finite impulse response  filter is used to represent channel impulse response $\bf h$ for generating the short channel. Each tap is modeled as an independent complex Gaussian random variable and set at integer multiples of the sampling interval. The maximum number of paths is decided by $\tau_{rms}$ and sampling period $T_{s}$. In this study, the maximum number of paths $n_{max}$ is set as  $\frac{10\tau_{rms}}{T_{s}}$.  This model is called EXP($n_{max}$) channel model in the following part.
	
	\textbf{Long channel}
	uses the Stanford University Interim (SUI) channel model [11],  {which is a type of outdoor multipath channels}. In IEEE 802.16, the suburban path loss environment can be divided into three terrains according to the tree densities and path-loss conditions. This channel can be described by different combinations of channel parameters. For SUI-5 channel model,  the delay spread is $[0~0.4n_{\max} ~n_{\max}]$ and power profile is $[0~dB~-5~dB~-10~dB]$, where
	\begin{equation}
	n_{\max}=\lceil \frac{10\tau_{rms}}{T_{s}} \rceil.
	\end{equation}
  In the following part, SUI-5 environment with the max delay spread at $n_{\max}$ is denoted as SUI-5($n_{\max}$).

	\textbf{Robust channel} is based on the assumptions in \cite{701317}.  The delay spread is $[0~1 ~\cdots ~n_{\max}]$ and the power profile is $[\frac{1}{n_{\max}+1}~\frac{1}{n_{\max}+1} ~\cdots ~\frac{1}{n_{\max}+1}]$.
	
	\subsubsection{Conventional LMMSE receivers}The channel correlation matrix and the noise power are the key parameters for the LMMSE receivers. The noise power can be easily estimated from the subcarriers for guard. However, the channel correlation matrix is  challenging  because of the varying channel environment. 
	
	\textbf{Online LMMSE} is based on the method of estimating delay spread in  \cite{hung2010pilot}. It assumes  the channel is with multipath fading and its PDP is with exponential shape similar to Short Channel. Therefore, the element in the channel autocorrelation matrix  can be expressed as
	\begin{equation}
	R_{f}(k)/R_{f}(0)= \frac{e^{-j2 \pi \tau_{0} k/N}}{1+j2 \pi \tau_{rms} k/N},
	\end{equation}
	where  $\tau_{\mu}$ denotes mean delay,  $\tau_{0} = \tau_{\mu} - \tau_{rms}$; $N$ is the size of the DFT used in OFDM modulation.
	
	Then,  the LS CE and DD (decision-directed) CE are used to estimate  $\tau_{rms}$ and $\tau_{\mu}$. For example, if the pilots are limited, then the known data blocks are used to obtain channel information.  We use the same training frames  for Online LMMSE  and the SwitchNet for comparison.
	
	\textbf{Robust LMMSE} is based on the work in [1], which calculates the max delay and the LMMSE matrix offline. Therefore,  the channel autocorrelation matrix  can be expressed as
	\begin{equation}
	\bf{R}_{f}= \bf{FDF}^{H},
	\end{equation}
	where $\bf F$ is discrete DFT matrix and $(\cdot)^H$ is Hermite transpose. $\bf D$ is diagonal matrix with the elements  $diag\{\frac{1}{n_{\max}+1},\frac{1}{n_{\max}+1},\cdots,\frac{1}{n_{\max}+1},0,\cdots,0\}$. In the following part, the robust LMMSE receiver designed with the max delay of $n_{\max}$ is called Robust LMMSE $n_{\max}$.

	 In this study,  Perfect baseline has knowledge of true channel and represents the best performance of a linear receiver. The statistic parameters of the channel are changing and difficult to obtain. Thus, the Robust LMMSE and the Online LMMSE are also simulated for comparison. The performance of the two LMMSE estimator is determined by the accuracy of statistical information of channels. The noise is set as 40 dB, which is the same as the training environment of NNs.

	\subsubsection{Parameter setting}
	The detailed network layouts of the AI-aided OFDM receivers are summarized in Table \ref{AI receivers}. Training parameters are shown in Table \ref{train}. The parameters in the AI-aided OFDM receivers are trained through labeled data in advance. Table \ref{train} lists the selected training parameters in the simulation.

	\begin{table}[!h]
		\centering	
		\caption{Network layouts of the AI-aided OFDM receivers. }
		\footnotesize
		\begin{tabular}{>{\sf }c|c|c|c|c}    %
			\toprule
			& & Layer & Output & Activation\\
			& & & dimensions & function  \\ \hline
			\multirow{5}{*}{FC-DNN}	&&Input&	 256 & None \\
			& & FC &	 500& ReLU \\
			& & FC &	 250& ReLU \\
			& & FC &	 120& ReLU \\
			& & FC &	 16& Sigmoid \\ \hline
			\multirow{5}{*}{ComNet}  &\multirow{2}{*}{CE}&LS Estimation &	 128 & / \\
			& & FC &	 128& None \\ \cline{2-5}
			&\multirow{3}{*}{SD}
			& ZF Detection & 128 & / \\
			& & FC &	 120& ReLU \\
			& & FC &	 16& Sigmoid \\ \hline
			\multirow{7}{*}{SwitchNet}&\multirow{3}{*}{CE}&LS Estimation &	 128 & / \\
			&	& FC1 &	 128& None \\
			
			&	& FC2&	 128& None \\ \cline{2-5}
			&  &  FC1 out + FC2 out& 	128 &/ \\ \cline{2-5}
			&\multirow{3}{*}{SD}
			& ZF Detection & 128 & / \\
			& & FC &	 120& ReLU \\
			& & FC &	 16& Sigmoid \\ \bottomrule
		\end{tabular}
		\label{AI receivers}
	\end{table}
	
	\begin{table}[!h]
		\centering
		\caption{Training parameters in simulation.}	
		\footnotesize
		\begin{tabular}{>{\sf }c|c}    %
			\toprule
			Parameter&   Value   \\
			\hline
			SNR&   25 dB   \\
			Loss function&	  MSE \\
			Epoch&	 2000 \\
			Initial learning rate&	  0.001 \\
			Optimizer&	  Adam \\
			\bottomrule
		\end{tabular}
		\label{train}
	\end{table}
	
		\subsection{Performance of the existing AI-aided OFDM receivers}

The existing AI-aided OFDM receivers, namely, the FC-DNN and ComNet, adopt offline training but an online deployment scheme. We will evaluate the performance variation of FC-DNN and ComNet when they encounter mismatched channels. The Robust LMMSE, Online MMSE and Perfect baseline are tested for comparison.

\begin{figure}[!t]
		\centering
		\subfloat[ ]{
			\includegraphics[width=3in]{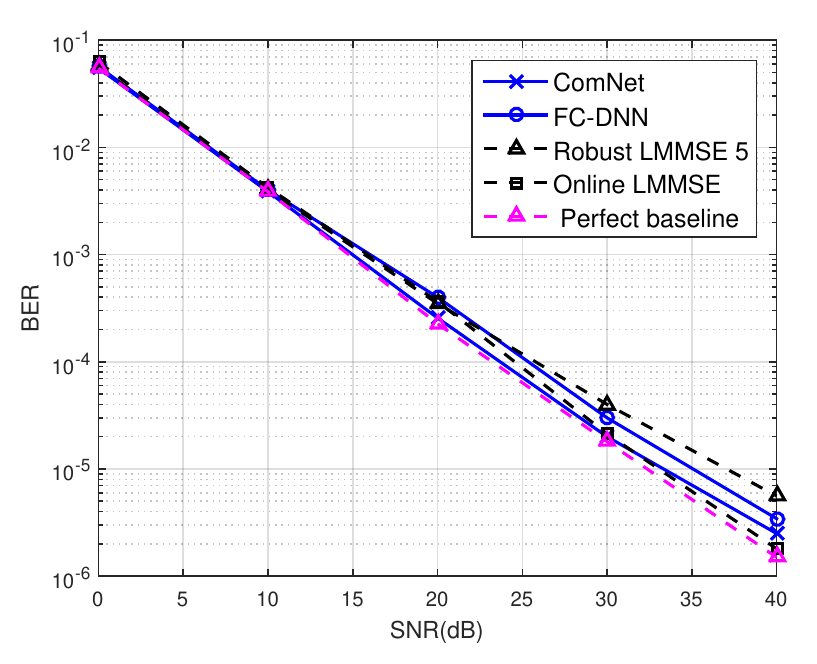}}
		\subfloat[ ]{
			\includegraphics[width=3in]{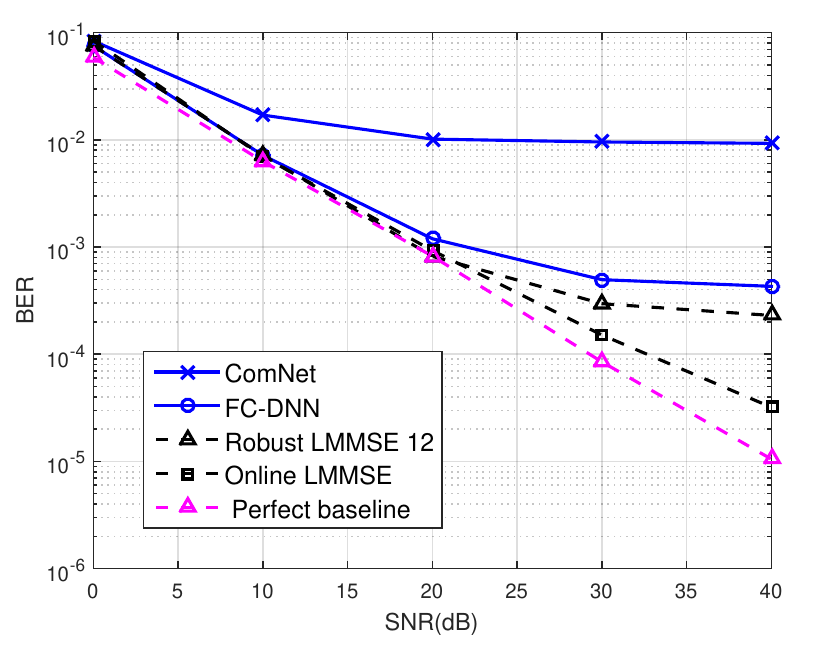}}
		\caption{BER performance of FC-DNN and ComNet under matched  and mismatched channels. (a)  FC-DNN and ComNet receivers  trained and tested under matched channels. (b)  FC-DNN and ComNet receivers  trained and tested under mismatched channels.}
		\label{Channelerr}
	\end{figure}

	Fig. \ref{Channelerr}(a) compares the BER performance of ComNet, FC-DNN, and LMMSE. The AI-aided methods are trained and tested in EXP(5) channel, which means the trained and  tested channels are the same. Fig. \ref{Channelerr}(a) shows that these  receivers exhibit similar BER performance when SNR $\leq$ 10 dB given that the influence of the accurate delay spread is slight when the noise power is high.  The ComNet is close to the Perfect baseline and nearly 1 dB better than the Online LMMSE when SNR $=$ 20dB. With the increase in SNR, the ComNet is slightly worse than Online LMMSE. This result  means that the ComNet learns more to deal with the noise. Meanwhile, the proper channel shape assumption of the Online LMMSE results in its tightly close performance  to the Perfect baseline, especially when SNR is high.  The evident performance gain of ComNet over FC-DNN suggests that the expert knowledge of the traditional algorithm can be beneficial to the learning process of the DL networks. The Robust LMMSE shows the worst performance because it uses the least statistic information.

Fig. 7(b) compares the BER performance of ComNet and FC-DNN  {trained in the EXP(5) channel but} tested in the SUI-5(10) channel. As shown in the figure, the ComNet is the worst. The BER performance of the FC-DNN receiver is still close to that of Robust LMMSE, whereas the ComNet receiver does not work well and becomes saturated when SNR $>$ 20 dB. The Online LMMSE is still close to the Perfect baseline because it can learn the changing channel and recalculate the LMMSE matrix. However, a gap exists between the Online LMMSE and the Perfect baseline when the SNR is high because the exponential model assumption is inaccurate. In general, the ComNet receiver has excellent performance under matched channels and poor performance under mismatched channels. By contrast, the FC-DNN is more robust to channel mismatch than ComNet, which may be due to the redundant network parameters. The Robust LMMSE is worse and shows error floor because its performance degrades when the delay spread is longer. 

The existing AI-aided OFDM receivers outperform the traditional method, especially under the non-ideal scenarios, such as insufficient pilots, clipping, and other nonlinear effects \cite{8052521}. However, the feasibility of the AI-aided method is still questionable. 
	


	\subsection{Performance of the SwitchNet receiver}
	
Performance degradation of the existing AI receivers for mismatched channels is due to their offline training mode, which makes them well known to the trained channel  but ``unfamiliar'' with the untrained channels. Performance may not be guaranteed for the AI receivers under
real scenarios with channels untrained offline. The AI receiver should be trained under more channel models offline or the online receiver should be trained to adapt to the real environment, as indicated in the proposed SwitchNet, to address the channel mismatch issue.

\begin{figure}[!h]
		\centering
		\subfloat[ ]{
			\includegraphics[width=3in]{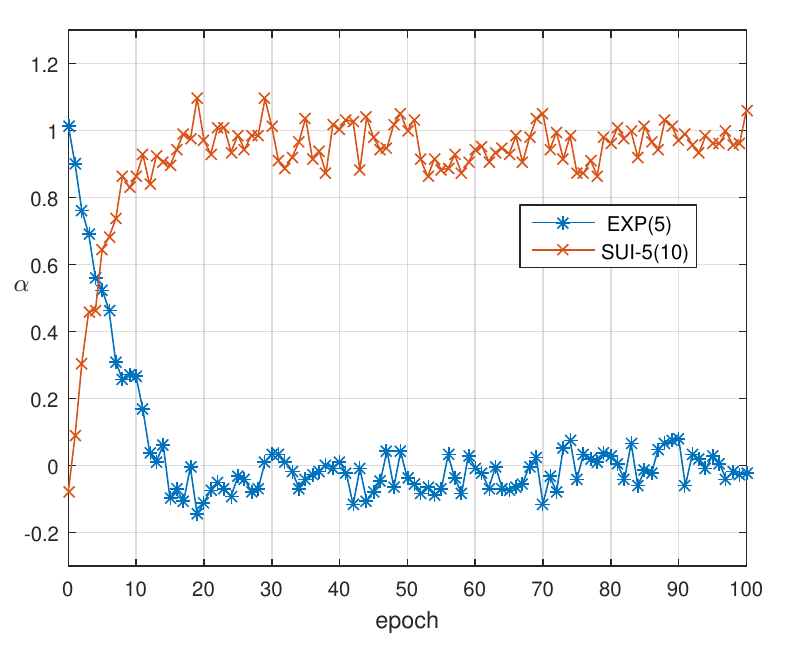}}
		\subfloat[ ]{
			\includegraphics[width=3in]{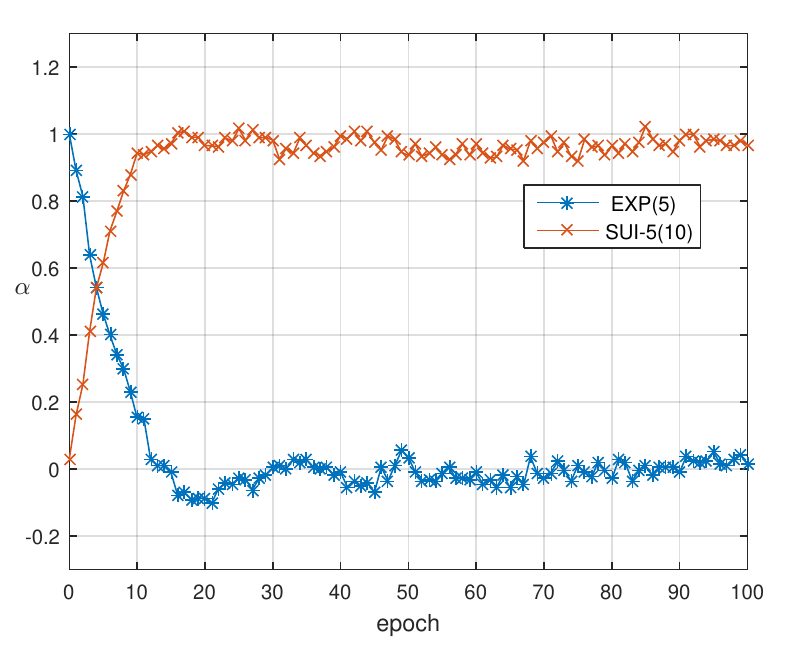}}
		
		\caption{Online training process during channel change. The star curve is the training process of $\alpha_{1}$ when the channel changes from SUI-5(10) to EXP(5). The crossing curve is the training process of $\alpha_{1}$ when the channel changes from EXP(5) to SUI-5(10).}
		\label{Swichpara}
	\end{figure}

 We compare the different sizes of online training set to verify the feasibility of the SwitchNet. The SwitchNet is combined with two CE RefineNets trained under EXP(5) and SUI-5(10), respectively. We let $\bm{\alpha}=[1~0]$ to train CE RefineNet 0 for EXP(5) and $\bm{\alpha}=[1~1]$ to train CE RefineNet 1. Fig. \ref{Swichpara}(a) uses only one subframe for online training, and Fig. \ref{Swichpara}(a) uses 10 subframes. The learning rate is set as 0.1, and 50 times of training are conducted  for an epoch. 
	
Before training online, the receiver works in the specific channel environment. The value of $\alpha_{1}$ is 0 when the simulated environment is EXP(5) or $\alpha_{1}$ is 1 when the environment is SUI-5(10). When the channel suddenly changes, the value of $\alpha$ adjusts immediately to match the new channel. Fig. \ref{Swichpara} shows the online-training curves when the channel changes. As shown in the figure, the value of $\alpha_{1}$  of the   star curve changes quickly from 1 to 0 within 10 epochs when the channel changes from SUI-5(10) to EXP(5). Similarly, the { crossing} curve adapts to 0 from 1 within 10 epochs when the channel changes from EXP(5) to SUI-5(10). Within 10 epochs,   $\alpha_{1}$ moves closer to the value of 0 or 1 and slightly varies around them. The amplitude of variation decreases gradually and converges eventually.  Although the vibration of the curve in Fig. \ref{Swichpara}(a) is larger than  that in Fig. \ref{Swichpara}(b), the BER performance is still tightly close to that of the ComNet trained under the matched channels. Therefore, the online system can perform well in terms of adaptability and stability. In our simulation, the 100 epochs can be trained within 1 ms.

A robust combination of the CE RefineNets is proposed to explain the expandable architecture of SwitchNet. The CE RefineNet 0 is initiated by the Robust LMMSE 12 and trained for better performance.   After CE RefineNet 0 is trained, we set ${\bm \alpha}=[0~1~0~0~0]$ and train for the max delay at 3. Then, we set $\bm{\alpha}=[0~1~1~0~0]$ for the max delay at 6, $\bm{\alpha}=[0~1~1~1~0]$ for the max delay at 9, and $[0~1~1~1~1]$ for the max delay at 12.  The pilot numbers are set as 8 and 64 for comparison in this simulation.

	\begin{figure}[!ht]
		\centering
		\subfloat[ ]{
			\includegraphics[width=3in]{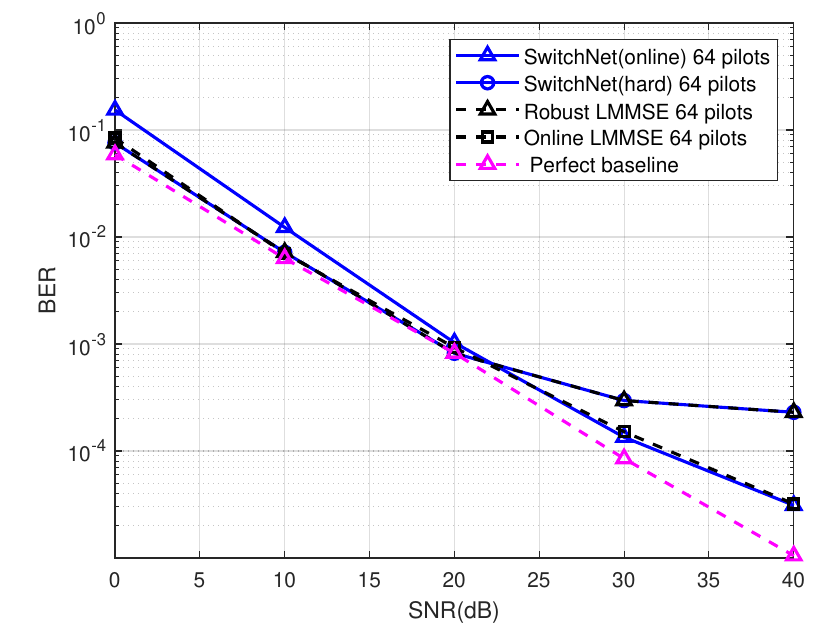}}
		\subfloat[ ]{
			\includegraphics[width=3in]{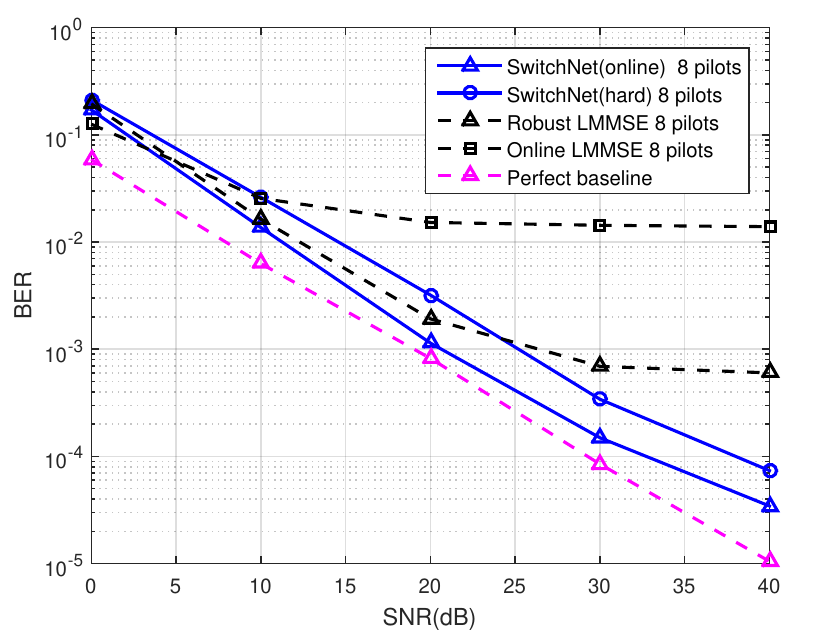}}
	\caption{BER performance of SwitchNet after online training when the channel changes. The test scenario is SUI-5(10), where the SwitchNet has not learned offline.}
		\label{SwitchBER}
	\end{figure}

	Fig. \ref{SwitchBER} shows the BER performance of the SwitchNet receiver after online training from the initial state, where $\bm{\alpha}=[1~0~0~0~0]$. We use 10 OFDM symbols as a training set and train 50 times for an epoch. All the online training process converges within 40 epochs.  The SwitchNet(online) is learned by gradient descent as in Fig. \ref{Swichpara} and the SwitchNet(hard) means all the $\bm{\alpha}=$ values are chosen either 0 or 1 by comparing the BER performance. Thus, if we train several ComNets offline and choose the best online, then the ComNet has the same BER performance as the SwitchNet(hard).  $\bm{\alpha}=[0.25~0.73~0.52~0.04~0.00]$ for SwicthNet(online) with 64 pilots,  and $\bm{\alpha}=[0.13~0.86~0.65~ 0.09~0]$ for SwitchNet(online) with 8
	pilots. The SwitchNet(hard) chooses the best combination in this scenario and when $\bm{\alpha}=[0~1~1~1~0]$. We find that the SwitchNet(online) uses the CE RefineNets from 0 to 3 because the CE RefineNet 0 is robust and the CE RefineNets from 1 to 3 are trained when $n_{max}=9$. This online training result is reasonable because the $n_{max}$ of SUI-5(10) is 10.

	Fig. \ref{SwitchBER}(a) shows that the SwitchNet online is close to Online LMMSE and the SwitchNet hard is close to Robust LMMSE when SNR is high. The two online methods are better where the robust methods offline have error floor. This performance is due to the aid of online training. Online LMMSE is better than the SwitchNet online when SNR $\leq$ 20 dB. The performance loss is reasonable because Online LMMSE has an exponential channel shape assumption, which only has a slight model error. The SwitchNet(online) needs to combine the  CE RefineNets properly online while it has only learned the Robust channel.  When the pilot number reduced to 8 in Fig. \ref{SwitchBER}(b), the superiority of the SwitchNet  to the conventional ways is obvious.  Meanwhile, the SwitchNet(online) is still better than the SwitchNet(hard). Therefore, the several trainable parameters for online training enable the AI-aided method to adapt to the unfamiliar scenarios.  Robust LMMSE can still work due to its robust design but  Online LMMSE fails. 
	
	\begin{figure}[!ht]
	\centering
	\subfloat[ ]{
		\includegraphics[width=3in]{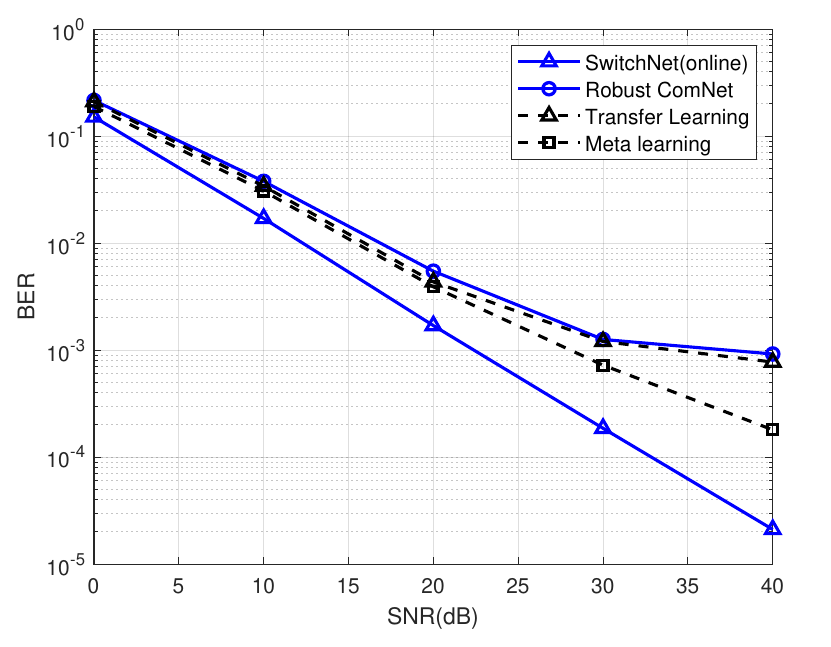}}
	\subfloat[ ]{
		\includegraphics[width=3in]{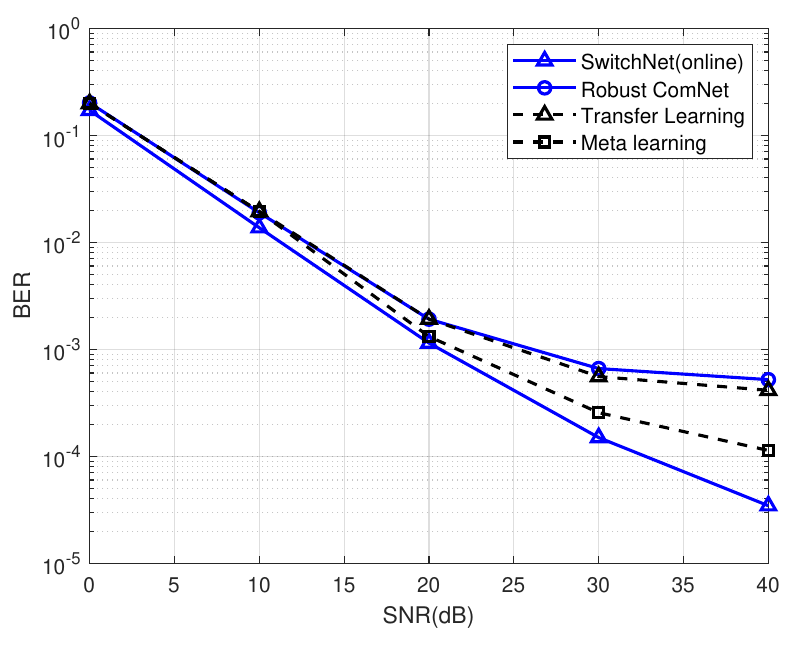}}
	\caption{BER performance of SwitchNet and competing methods with 8 pilots. (a) Under the  scenario of  CE RefineNet 2. (b) Under the SUI-5 (10).}
	\label{TM}
\end{figure}	

		We compare  different online training  methods with the SwitchNet in Fig. \ref{TM}. Robust ComNet is trained under all the possible scenarios offline, which is the common method to cope with the changing scenarios. Transfer learning and meta-learning  \cite{finn2017model} use  Robust ComNet as initiation. A total of 10 collected  subframes are used to fine tune the CE subnet given a fixed SD subset. The Robust ComNet can work under all  scenarios but cannot achieve the best performance, which shows the similar trend to the Robust LMMSE in Fig. \ref{SwitchBER}. The transfer learning method only has a slight improvement over the Robust ComNet because  10 subframes are not enough to train an FC layer. Meta-learning is more suitable for such a few training data and better than the transfer learning. The SwitchNet shows the best performances and learns ${\bm{\alpha}}=[-0.01~0~0.98~0~0]$ in Fig. \ref{TM}(a) and $\bm{\alpha}=[0.13~0.86~0.65~ 0.09~0]$ in Fig. 9(b),  which demonstrates that 10 subframes are sufficient  for the SwicthNet to find a good combination of the CE RefineNets.

In summary, the online training process in the SwitchNet receiver can combat performance degradation under the mismatching channel. Compared with training offline, the SwitchNet receiver needs much fewer training data and can work under varying channels OTA.  However, the CE RefineNets in SwitchNet should be trained under several possible environments, which are potentially suitable for practical channels. Otherwise, the performance will not be improved by online training. The combination of the CE RefineNets can rely on the expert knowledge in wireless communications to guarantee the robustness. The SwitchNet not only has the superiority of the AI-aided method under the non-ideal scenarios, such as no sufficient pilots, but also shows its expandability and flexibility for implementation.

	\subsection{Complexity analysis}	
	\begin{table}[h]
		\centering
		\caption{Forwarrd complexity analysis for SwitchNet and competing methods.}
		\label{Complexity}
		\footnotesize
		\begin{tabular}{>{\sf }lllll}    %
			\toprule
			&   FLOPs  &Activation memory & Parameters & Time \\
			\midrule
			SwitchNet  & 0.34M  & 10.50kBytes & 0.17M & 1.2us \\	
			ComNet    & 0.31M  & 9.47kBytes & 0.16M & 1.2us \\
			FC-DNN         & 4.33M  & 29.37kBytes & 2.29M & 1.2us \\
		{ 	LMMSE} & 0.06M & / & /& 0.8us \\
			\bottomrule
		\end{tabular}
	\end{table}	
	

Table \ref{Complexity} compares the complexity in terms of the number of floating-point multiplication-adds (FLOPs), activation memory consumption, parameters, and time consumption in one forward propagation to recover the binary bitstream in a frame among three AI-aided OFDM receivers. As shown in Table \ref{Complexity}, SwitchNet consumes slightly more resource than ComNet while remaining at a low complexity compared with FC-DNN. Specifically, SwitchNet needs 0.03 million more FLOPs, 1.03 thousand more bytes activation memory, and 0.01 million more parameters than ComNet while only costing approximately 1/10 of hardware resources compared with FC-DNN. The additional hardware consumption of SwitchNet relative to that of ComNet is reasonable. SwitchNet, which is an enhanced architecture of ComNet, has an extra CE subnet to adapt to added channel models; thus, it has a slightly larger hardware consumption than ComNet. Meanwhile, the running time of the three AI-aided OFDM receivers is comparative due to paralleled calculation of graphic processing unit (GPU) and the same depth of network.  The traditional OFDM receiver in our system has an LMMSE estimator and an MMSE detector, and its complexity is always the lowest.  The FLOPs of a CE subnet is similar to that of the LMMSE estimator, but that of the SD subnet is much larger because of the fully connected architecture and only one-eighth of the transmitted bits demodulated once. Thus, the FLOPs is 8 times that in \cite{gao2018comnet}. The high complexity of SD subnet can be reduced  if the experimental scenarios are simple. By clipping the size of hidden layer and removing the nonlinear activation function, the FLOPs of the SwitchNet is close to 0.1 M. The reasons for simplifying the SD subnet are discussed in Section IV D.

Overall, complexity analysis suggests that SwitchNet has the advantage of adaptability to more channel models with acceptable sacrifice in hardware resource compared with ComNet. Moreover, SwitchNet consumes considerably fewer hardware resources than FC-DNN.

	\setlength{\baselineskip}{22pt}
	\section{OTA Test and Discussion}
	

 OTA test is  important for AI-based receivers because  considering all distortions in the systems is nearly impossible. Several prototyping systems have been developed to verify the feasibility and effectiveness of AI-based receivers in real environments. In \cite{yang2017rapro}, a novel 5G RaPro system is proposed to deploy FPGA-privileged modules on SDR platforms, implement complex algorithms on multi-core GPPs, and connect them through high-speed 10-gigabit Ethernet interfaces. Such architecture deploys a multi-user full-dimension MIMO prototyping system \cite{yang2017rapro, Gao2018Implementation} and  is therefore flexible and scalable. In this study, we set up the first real-time testbed for AI-aided OFDM receivers, which has been never reported before. {The RaPro system is used as our testbed to test the OTA performance of the FC-DNN, the ComNet, and the SwitchNet receivers.} Various tests are conducted in different scenarios. After  the system setup and software implementation are introduced, we will present our experimental results and analyze the flexibility of the AI-based receiver.
	
	\subsection{System setup}

\begin{figure*}
		\centering
		\includegraphics[width = 4in]{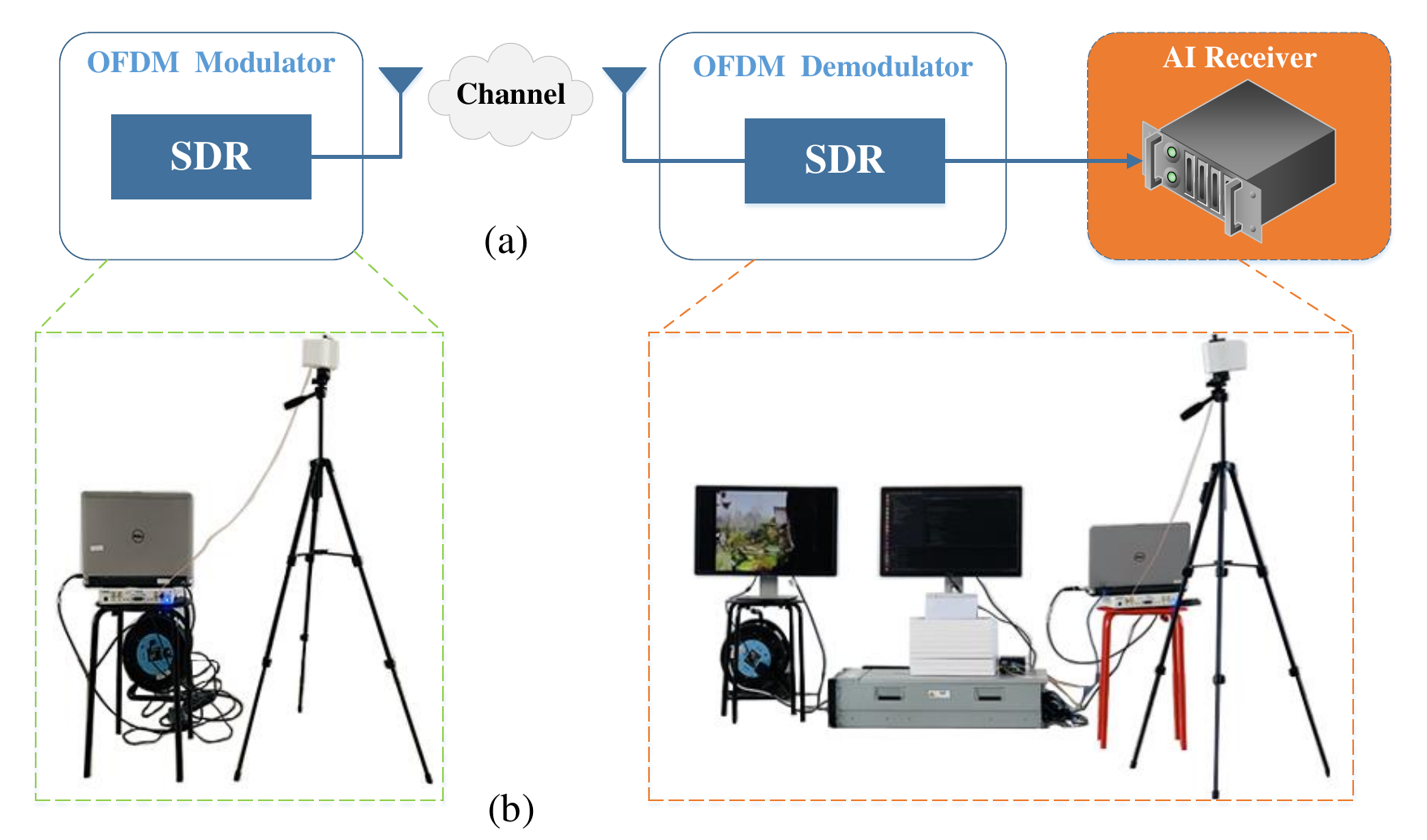}
		\caption{AI-aided OFDM receiver system based on the RaPro architecture. OFDM-related algorithms are deployed on SDRs while AI receivers are deployed on the multi-core server.}
		\label{hardware}
	\end{figure*}

Fig. \ref{hardware}(a) illustrates the AI-aided OFDM receiver based on the RaPro architecture. It comprises  SDR nodes  at the transmitter and the receiver, respectively, and a multi-core server. Each SDR has with one antenna and one RF chain and is provided with a unified reference clock and trigger signal by the timing/synchronization module. The AI-aided OFDM receivers are implemented on a multi-core server in a Linux environment. The proposed receivers (FC-DNN, ComNet, and SwitchNet) can be developed on multi-core GPPs by programming with high-level language, such as C/C++, in conjunction with Intel Math Kernel Library (MKL), which is a highly optimized and commonly used math library for processors.
	
Fig. \ref{hardware}(b) shows the hardware of the assembled AI-aided OFDM receiver system, which is implemented by two SDR nodes of USRP-2943R and a multi-core server that contains 32 Intel Xeon E5-2680 v2 @ 2.8 GHz processors. Each SDR node consists of two RF transceivers of 120 MHz bandwidth, one of which is used to transmit modulated radio signals. The multi-core server provide sufficient GPPs to meet the requirements of TensorFlow and MKL, which are necessary for the implementation of the AI-aided receivers. An RF antenna of USRP-RIO,  which has  center frequency is adjustable from 1.2 GHz to 6 GHz,  receives the wireless signals.  After CP removal and FFT-based OFDM demodulation operated by USRP-RIO, the data are sent to the multi-core server via a cable. The video stream is recovered by the AI receiver running on the server.

{  The SwitchNet receiver can be easily extended to a MIMO system because the SwitchNet receiver is inherited from the ComNet, in which the traditional methods, including LS CE and ZF SD, can be used.}
	
	\subsection{Software implementation}
	
On the transmitter side, the video stream is transmitted through the RF module after QPSK modulation and IFFT. On the receiver side, FFT transformation is performed to the signal received by the antenna. Then, the data are sent to the multi-core server through user datagram protocol (UDP) module. The AI-aided OFDM receivers (FC-DNN, ComNet, and SwitchNet), which run on the multi-core server, will recover and display the original video stream.

\begin{figure*}
		\centering
        \includegraphics[width = 4in]{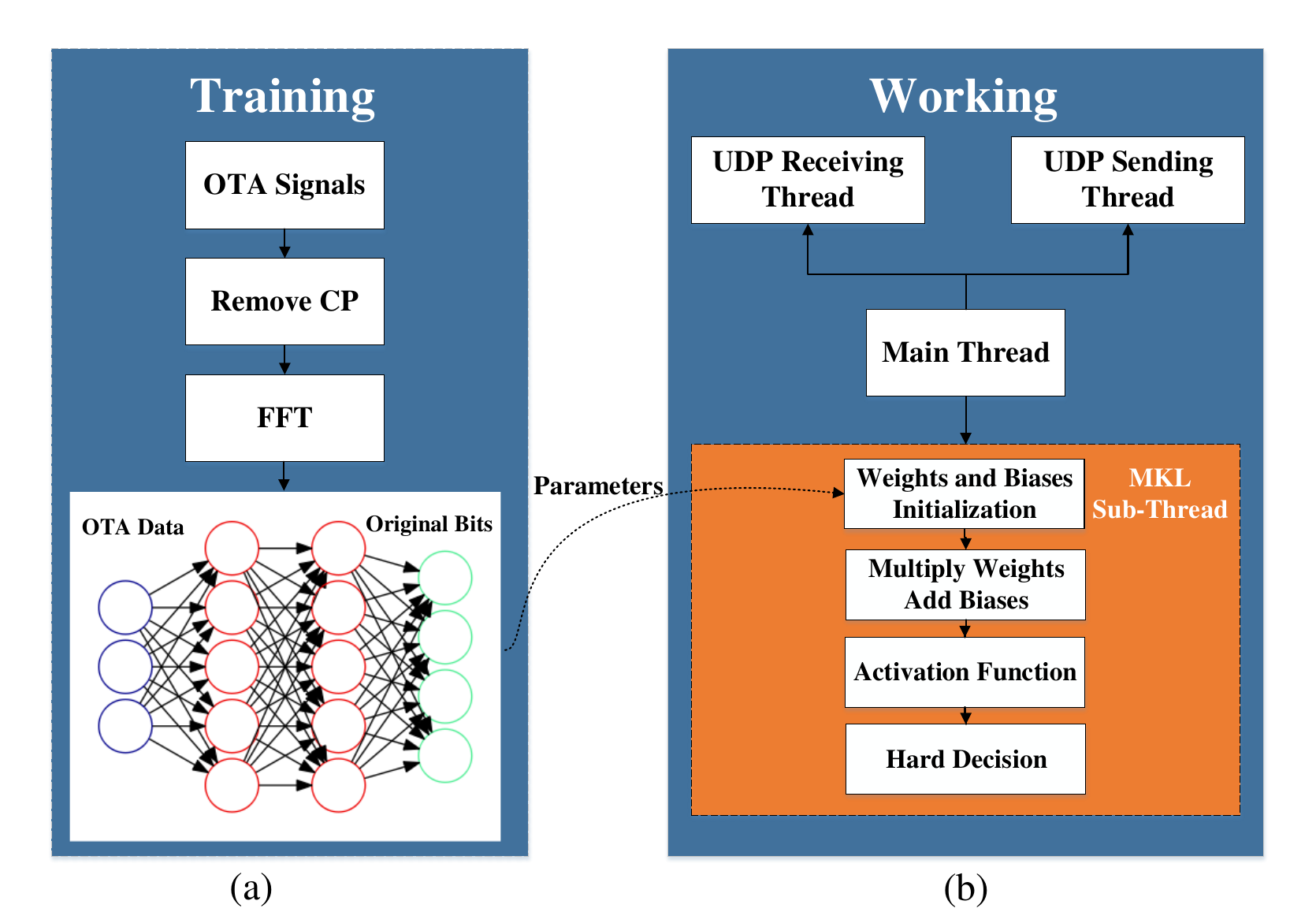}
		\caption{Overall data processing program of training and working phases. The weights and biases of the AI receiver  trained based on TensorFlow and will be used to initialize the parameters of the matrices in the working phase.}
		\label{WorkingAndTraining}
	\end{figure*}

The proposed AI-aided OFDM receiver contain two phases: training and working. The training phase is realized in Python based on TensorFlow and  powerful GPU.  The back propagation algorithm is used to train the weights and biases of the DNN via OTA data captured by USRP-RIO. These parameters are saved as comma-separated value (csv) files after training and used for the working phase. In the working phase, the values of the weight matrices and bias vectors are initiated by the parameters in the csv files. Then, the forward propagation is performed in C/C++ with the help of Intel MKL library on the multi-core server.  Fig. \ref{WorkingAndTraining}(a) shows the architecture of the training phase. After the zero padding removal module, 128 effective subcarriers of pilot and data are saved. A total of 256 real inputs are prepared for FC-DNN through the separation of their real and imaginary parts. For ComNet, the received pilot divides the local pilot to obtain LS CE. Similarly, the input of ComNet is real form of LS CE and data. Fig. \ref{WorkingAndTraining}(b) presents the overall data processing program diagram of the forward propagation on the multi-core server. Multi-threading technology is applied in the multi-core GPP-based AI-aided OFDM receiver design to process each module. Each processing thread is bounded to a unique central processing unit core with semaphore and spinlock as the synchronization mechanism to avoid the cost of context switching. A total of 11 threads are in the implemented system. The main thread is in charge of scheduling the other threads. A UDP receiving thread is used to collect the demodulated data from USRP-RIO. Eight AI detection (FC-DNN and ComNet) threads run in parallel, where the matrix manipulation in the forward propagation is realized based on the Intel MKL Library. After detection, one UDP transmitting thread is used to pack the video stream and send to display. 

{  Considering that the data transmission and preprocessing capabilities of the receiver (e.g., receiving data, unpacking data, and repacking data to transmit to server) are limited by the host of the USRP, the testbed adopts 15 ms frame duration and 731 kHz sample rate. According to the frame structure in Fig. \ref{frame}, each frame contains $36\times64\times2=4,608$ bits, where 36 represent the number of subframes to transmit valid data and pilots, 64 are for 64 subcarriers to transmit valid modulated symbols, and 2 are for 2 bits in a QPSK symbol. Therefore, the data transmission rate is 4608 bits/15 ms$=307$ kbps. This data transmission rate can be increased if the data transmission and preprocessing strategy of the receiver can be accelerated by FPGA implementation.}

\subsection{Implementation details}
	\label{Implementation_details}
	
	\subsubsection{OTA scenarios for offline-trained AI receivers}
	\label{scenarios}
	
	\begin{figure}[!h]
		\centering
		\includegraphics[width=3.5in]{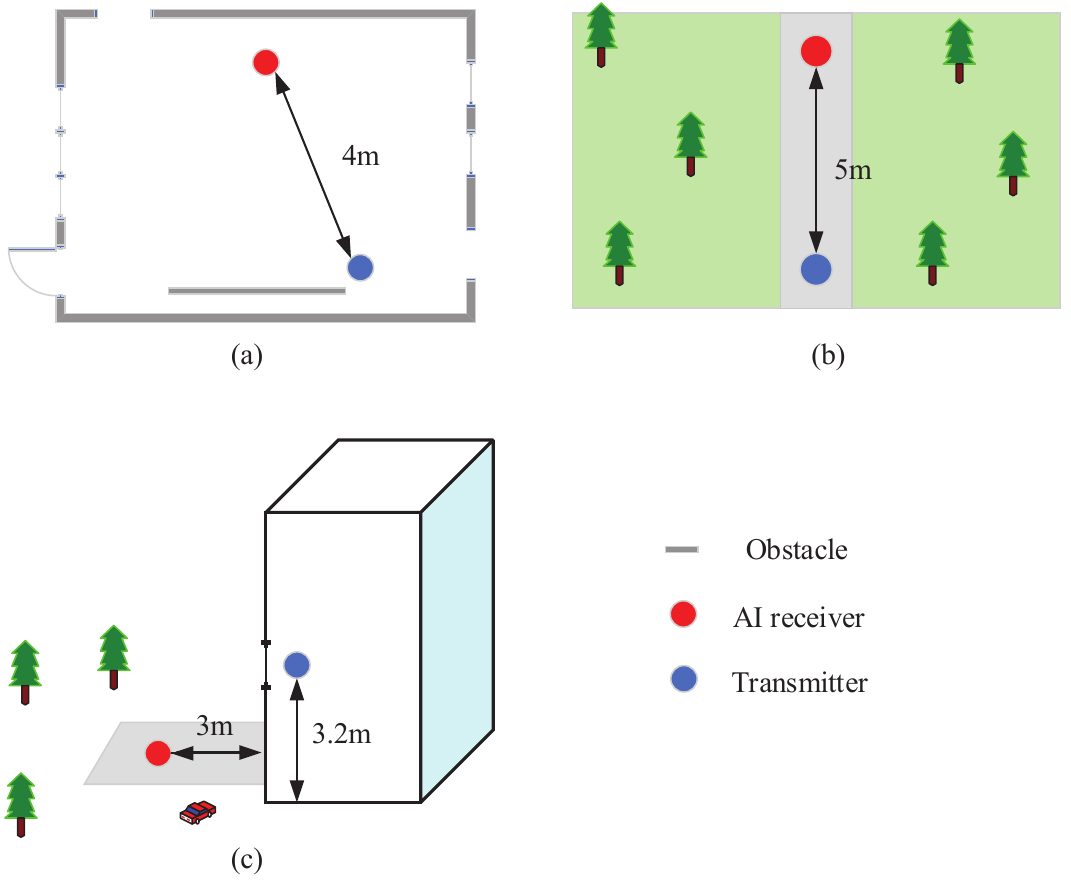}
		\caption{Three scenarios of the OTA test for real-time AI testbed: (a) indoor scenario with an obstacle and windows, doors, and walls  around; (b) outdoor scenario where the transmitter and receiver are placed on a straight road surrounded by trees and grass; and (c) indoor to outdoor scenario in which the transmitter is deployed on the second floor of the building and the receiver is outside the building surrounded by several trees and cars.}
		\label{scenario}
	\end{figure}

As  shown in Fig. \ref{scenario},  three different scenarios are chosen to test our real-time AI testbed.
\begin{itemize}
	\item {Scenario 1 is the indoor scenario in Fig. \ref{scenario}(a), where the transmitter and  the receiver are in the same room with obstacles, windows, and walls around; the distance between the transmitter and the receiver is 4 m.}
	\item Scenario 2 is the outdoor scenario in Fig. \ref{scenario}(b), where the transmitter is at a distance of 5 m on a straight road surrounded by several trees.
	\item In Scenario 3, as shown in Fig. \ref{scenario}(c), the transmitter is deployed indoors while the AI receiver is deployed outside the building.
\end{itemize}

     The practical channels can be regarded as the EXP channel model because of
     the limited transmission distance, reflectors, and scatters of the three scenarios. Therefore, the EXP channel model is applied to train the FC-DNN, ComNet, and SwitchNet receivers offline  and  conduct the OTA test under high and low SNR by changing the antenna gain of the testbed.
	
	\subsubsection{Training strategy for online-trained AI receivers}

	In the real-time system, the AI receivers obtain online training dataset by the received training sequence that is sent by the transmitter and known to the receiver. We use pseudo random testing dataset and calculate BER  to measure the online training performance of the AI receivers. In \cite{gao2018comnet}, the CE subnet is trained independently, which is very difficult in online training because the accurate information of the real channel remains unknown. Thus, the parameters of ComNet are refined by the online training dataset in an end-to-end manner, which is the same as the FC-DNN receiver. The online training method of FC-DNN and ComNet corresponds to the idea of transfer learning.  We use the frame structure depicted in Fig. \ref{frame}, that is, one pilot symbol followed by one data symbol, for real-time transmission. The training process is similar to that in Fig. \ref{Swichpara}(b).

	
	\subsection{OTA performance of offline-trained AI receivers}
	In this subsection, we compare the experimental results of the offline-trained FC-DNN and ComNet receivers in the OTA tests.
	

The two receivers are trained offline under the EXP(5) channel model. The Online LMMSE is used as the baseline. As shown in Table \ref{table_ber}, the LMMSE method achieves better BER performance than the two AI-aided OFDM receivers in all scenarios and the FC-DNN receiver slightly outperforms the ComNet receiver. The mismatch phenomenon is observed. 

	\begin{table}[!h]
		\centering	
		\footnotesize
		\caption{BER performance of AI receivers and the LMMSE receiver in OTA test(mismatch)}
		\begin{tabular}{>{\sf }ccrrr}    %
			\toprule
			& SNR &  LMMSE  & FC-DNN & ComNet  \\
			\midrule
			\multirow{2}{*}{Scenario 1}&High SNR   &  $\bf{1.74\times 10^{-6}}$ &${5.21\times 10^{-6}}$   & ${5.21\times 10^{-6}}$ \\	
			&Low SNR    & $\bf{1.88\times 10^{-4}}$ &${3.68\times 10^{-4}}$& ${3.94\times 10^{-4}}$   \\
			\hline
			\multirow{2}{*}{Scenario 2}	&High SNR   &$\bf{5.99\times 10^{-5}}$ & ${1.10\times 10^{-4}}$  & ${1.11\times 10^{-4}}$ \\
			&Low SNR    & $\bf{4.71\times 10^{-4}}$ & ${7.36\times 10^{-4}}$&${7.73\times 10^{-4}}$  \\
			\hline
			\multirow{2}{*}{Scenario 3}	&High SNR   & $\bf{2.78\times 10^{-5}}$ &${5.82\times 10^{-5}}$  & ${7.52\times 10^{-5}}$ \\
			&Low SNR    & $\bf{1.30\times 10^{-5}}$ & ${2.86\times 10^{-5}}$ &${5.29\times 10^{-5}}$   \\
			\bottomrule
		\end{tabular}
			
		\label{table_ber}
	\end{table}

The offline OTA data from the SISO system with 153.6 MHz sampling rate are used for test to investigate the aforementioned issue, and Table \ref{SD} shows the  result. The Online LMMSE is better than FC-DNN in Scenario 1 but worse in others. The ComNet shows the best performance in all scenarios, and this performance  verifies that the channel realization in our online video transmission system is simple due to low sampling rate. We also find that the nonlinear activation should be removed  because  the LS or LMMSE CE and ZF SD can perform well in the real-time video transmission. The nonlinear activation is unnecessary in simple channel condition and may worsen the BER performance because of its redundant complexity. 
	\begin{table}[!h]
		\centering	
		\caption{BER performance of AI receivers and the LMMSE receiver in OTA test (153.6MHz sampling rate)}
		\footnotesize
		\begin{tabular}{>{\sf }ccp{1.5cm}p{1.8cm}c}    %
			
		\toprule
		&  SNR & LMMSE  & FC-DNN & ComNet \\
		\midrule
		\multirow{2}{*}{Scenario 1}&High SNR       &${6.27\times 10^{-5}}$  &${8.89\times 10^{-5}}$  &$\bf{3.21\times 10^{-5}}$  \\ 	
		&Low SNR    &${3.91\times 10^{-4}}$  &${5.33\times 10^{-4}}$ &$\bf{1.72\times 10^{-4}}$ \\
		\hline
		\multirow{2}{*}{Scenario 2}	& High SNR    &${9.98\times 10^{-4}}$ &${7.84\times 10^{-4}}$  &$\bf{5.58\times 10^{-4}}$ \\
		&Low SNR    &${5.31\times 10^{-3}}$  &${4.44\times 10^{-3}}$  &$\bf{3.18\times 10^{-3}}$ \\
		\hline
		\multirow{2}{*}{Scenario 3}	&High SNR    &${6.60\times 10^{-4}}$ &${6.52\times 10^{-4}}$  &$\bf{5.37\times 10^{-4}}$ \\
		&Low SNR    &${5.74\times 10^{-3}}$&${4.29\times 10^{-3}}$ &$\bf{3.01\times 10^{-3}}$ \\
		\bottomrule
		\end{tabular}		
		\label{SD}
	\end{table}
	

	In summary, the expert knowledge in wireless communications is helpful when analyzing the effectiveness of the AI-aided methods. The simple channel realizations of our online video transmission system in the OTA test still lead to the performance gap between simulation and practice.

	\subsection{Online training for the AI receiver}
	\label{online_training}

In this subsection, we discuss online training for the AI-aided OFDM receiver.  Only the EXP(5) and SUI-5(10) are chosen to combat the channel environment with different lengths of channel spread for simplicity.  We perform transfer learning for ComNet and FC-DNN as in  Fig. \ref{TM} to demonstrate the superiority of the SwitchNet. In this case, the network is retrained  using online data in the transmission stage based on the offline-trained network.

	\begin{table}[!h]
		\centering
		\caption{The training process of $\alpha$ when initialized as one under the real channel.}	
		\footnotesize
		\begin{tabular}{>{\sf }crrrrr}    %
			\toprule
			epoch&0 & 10 &  20  & 50&100   \\
			\midrule
			$\alpha_{1}$	&1.0  & 0.107  & -0.168 & -0.065&-0.059    \\	
			\bottomrule
		\end{tabular}

		\label{onlineswitch}
	\end{table}			
	

Table \ref{onlineswitch} shows the change in $\alpha_{1}$ in the online training process, where the learning rate is optimized. The initial value of $\alpha_{1}$ is set to 1 as the network is initialized under the SUI-5 channel model and decreases to close to 0 within 20 epochs. This result indicates that SwitchNet can adapt to the simple real channel by online training data.     $\alpha_{1}$ is also convergent to -0.059, which implies that the real channel is not exactly EXP or SUI-5.  
%
%
\begin{table}[!ht]
		\centering
		\caption{{  BER performances of SwitchNet, ComNet and FC-DNN with different numbers of epochs and optimized learning rates.}}	
		\footnotesize
		\begin{tabular}{ccrrrr} %
			\toprule
			& & SwitchNet &  ComNet  & FC-DNN &LMMSE  \\
			\midrule
			\multirow{3}{*}{epoch}&{  0}  & ${2.2\times 10^{-2}}$   &${2.2\times 10^{-2}}$  & ${1.2\times 10^{-3}}$ & $\bf{4.7\times 10^{-4}}$  \\
			&10  & $\bf{4.7\times 10^{-4}}$  &${1.4\times 10^{-3}}$  & ${7.7\times 10^{-4}}$&  ${4.7\times 10^{-4}}$ \\	
			&100 &$\bf{4.5\times 10^{-4}}$  &${6.7\times 10^{-4}}$  & ${6.8\times 10^{-4}}$&  ${4.7\times 10^{-4}}$ \\
			\bottomrule
		\end{tabular}
		\label{transfer_learning}
	\end{table}


Table \ref{transfer_learning} compares the BERs of SwitchNet, ComNet, and FC-DNN by using online training with different numbers of epochs. The learning rate for each network is optimized. ComNet and FC-DNN are trained by transfer learning.  The table shows that the SwitchNet can perform online training rapidly with a few epochs. Meanwhile, ComNet and FC-DNN need  a relatively large number of epochs to yield a similar performance.
	
		\begin{table}[!h]
		\centering
		\caption{{  BER performances of SwitchNet, ComNet, and FC-DNN with different numbers of epochs and decayed learning rates.}}	
		\footnotesize
		\begin{tabular}{ccrrrr} %
			\toprule
			& & SwitchNet &  ComNet  & FC-DNN &LMMSE  \\
			\midrule
			\multirow{3}{*}{epoch}&{  0}  & ${2.2\times 10^{-2}}$   &${2.2\times 10^{-2}}$  & ${1.2\times 10^{-3}}$ & $\bf{4.7\times 10^{-4}}$  \\
			&10  & ${7.4\times 10^{-4}}$  &${1.1\times 10^{-2}}$  & ${1.4\times 10^{-3}}$&  $\bf{4.7\times 10^{-4}}$ \\	
			&100 &$\bf{4.5\times 10^{-4}}$  &${9.8\times 10^{-3}}$  & ${1.4\times 10^{-3}}$&  ${4.7\times 10^{-4}}$ \\

			\bottomrule
		\end{tabular}
		\label{learning_rate}
	\end{table}

%

We also investigate the effect of the learning rate for three networks.  Table \ref{learning_rate} shows that the initial learning rates for SwitchNet, ComNet, and FC-DNN are 0.6, 0.01, and 0.01, respectively. The learning rate is decreased to 1/5 when each 1/5 of the total epochs has been trained. Table \ref{learning_rate} illustrates that SwitchNet is relatively robust to the learning rate. Conversely, ComNet and FC-DNN heavily depend on the learning rate. An improper learning rate will result in severe deterioration, failure to restore performance through online training, and additional training data and time consumed.
	
	
{ The advantage of training with practical data has been verified by carefully tuning the parameters in the NNs. The bottleneck of training with practical data  is the convergence speed because the practical channel is varying with  time.}

{  Real-time transmission means the bitstreams are demodulated by the AI-aided receivers online in addition to capturing online data and testing offline.}   The real-time transmission is achieved only if the time consumption to process a frame on the receiver is less than the duration of a frame. After OTA test, the average duration of data processing on the receiver is 14 ms, which allows the realization of the real-time transmission. Among the processing modules, the average durations of the AI-aided OFDM receivers to conduct a forward-propagation inference are 2.4, 0.6, and 0.6 ms for FC-DNN, ComNet, and SwitchNet, respectively; they  account for 16\%, 4\%, and 4\% duty cycle   of a frame duration. The quick convergence of SwitchNet and ComNet compared  with that of FC-DNN is due to the introduction of expert knowledge in wireless communications to form the model-driven DL networks.

From the abovementioned discussion, we can conclude that SwitchNet is more promising than ComNet and FC-DNN in terms of online training. The SwitchNet can avoid overfitting and reduce training time because it needs only one
parameter to be optimized in the online training process. More trainable parameters can be introduced into the network to further improve the flexibility and adaptability given that the real-time system has adequate hardware resource and time for training model-driven AI networks.		
	
	\section{Conclusions and Future   Challenges}

We have proposed an online trainable AI-aided OFDM receiver, named SwitchNet, to adapt to the channel variation and diversity in the OTA   scenarios. The proposed SwitchNet receiver pretrains multiple channels offline and reserves an online trainable parameter to act as a switch that chooses the network for the real transmission. Simulation results indicate that the proposed SwitchNet receiver has feasibility in online training and outperforms the ComNet and FC-DNN receivers and the traditional LMMSE  baseline in terms of BER performance.  OTA tests  demonstrate BER gains under real scenarios and efficient online training characteristics of the proposed SwitchNet receiver.

{  Although the AI-aided OFDM receivers relieve the difficulty of mathematical modeling and may outperform conventional communication systems, a performance gap may occur between offline and the OTA test due to the mismatch between simulation and real environments. Considering all possible system imperfections is challenging. Online training is a promising method to solve this dilemma.  SwitchNet offers a realizable online training scheme by sharply reducing the number of parameters to be trained. The adaptive ability of SwitchNet is guaranteed by the addition of subnets that are trained offline under different channel models.}

	\bibliographystyle{IEEEtran}
	\bibliography{IEEEabrv,bibtex0320}
	
	%
	
	
	
	

\end{document}